\documentclass[fleqn,10pt]{SelfArx} 

\definecolor{color1}{RGB}{0,0,90} 
\definecolor{color2}{RGB}{0,20,20} 

\usepackage{hyperref} 
\hypersetup{hidelinks,colorlinks,breaklinks=true,urlcolor=color2,citecolor=color1,linkcolor=color1,bookmarksopen=false,pdftitle={Title},pdfauthor={Author},urlcolor=blue}

\usepackage{graphicx}
\usepackage[square]{natbib}
\usepackage{amsmath}
\usepackage{multirow}
\usepackage{subfigure}
\usepackage{pdflscape}
\usepackage{rotating}
\usepackage{caption}

\JournalInfo{Published in International Journal of Refrigeration, Vol. 33 (3), 437-448, 2010} 
\Archive{\href{http://dx.doi.org/10.1016/j.ijrefrig.2009.12.012}{DOI: 10.1016/j.ijrefrig.2009.12.012}} 

\PaperTitle{Review and comparison of magnet designs for magnetic refrigeration} 

\Authors{R. Bj\o{}rk, C. R. H. Bahl, A. Smith and N. Pryds} 
\affiliation{\textit{Department of Energy Conversion and Storage, Technical University of Denmark - DTU, Frederiksborgvej 399, DK-4000 Roskilde, Denmark}} 
\affiliation{*\textbf{Corresponding author}: rabj@dtu.dk} 

\Keywords{} 
\newcommand{\keywordname}{Keywords} 

\Abstract{One of the key issues in magnetic refrigeration is generating the magnetic field that the magnetocaloric material must be subjected to. The magnet constitutes a major part of the expense of a complete magnetic refrigeration system and a large effort should therefore be invested in improving the magnet design. A detailed analysis of the efficiency of different published permanent magnet designs used in magnetic refrigeration applications is presented in this paper. Each design is analyzed based on the generated magnetic flux density, the volume of the region where this flux is generated and the amount of magnet material used. This is done by characterizing each design by a figure of merit magnet design efficiency parameter, $\Lambda_\mathrm{cool}$. The designs are then compared and the best design found. Finally recommendations for designing the ideal magnet design are presented based on the analysis of the reviewed designs.}

\begin{document}

\flushbottom 

\maketitle 


\thispagestyle{empty} 

%

\section{Introduction}
Magnetic refrigeration is an evolving cooling technology that has the potential of high energy efficiency using environmentally friendly refrigerants. Magnetic refrigeration utilizes the  magnetocaloric effect (MCE), which is the temperature change that most magnetic materials exhibit when subjected to a changing magnetic field. This temperature change is called the adiabatic temperature change, $\Delta{}T_\mathrm{ad}$, and is a function of temperature and magnetic field. The temperature change is greatest near the Curie temperature, $T_\mathrm{c}$, which is different for different magnetocaloric materials \citep{Pecharsky_2006}. Because the MCE in the best magnetocaloric materials currently available exhibit a temperature change of no more than 4 K in a magnetic field of 1 T, a magnetic refrigeration device must utilize a regenerative process to produce a large enough temperature span to be useful for refrigeration purposes. The most utilized process for this is called active magnetic regeneration (AMR).

At present, a great number of magnetic refrigeration test devices have been built and examined in some detail, with focus on the produced temperature span and cooling power of the devices \citep{Barclay_1988, Yu_2003,Gschneidner_2008}. So far the magnet, a key component in the magnetic refrigeration system, has been largely overlooked, even though it is often the single most expensive part of a magnetic refrigerator. Also little effort has been made to compare existing magnet designs in order to learn to design more efficient magnetic structures.

In general a magnet design that generates a high magnetic flux density over as large a volume as possible while using a minimum amount of magnet material is to be preferred. Since the magnet is expensive it is also important that the magnetic refrigerator itself is designed to continuously utilize the magnetic flux density generated by the magnet.


\subsection{Magnetic refrigeration magnets}
As previously stated a substantial number of magnetic refrigeration devices have been built. In all devices, one of three types of magnets has been used to generate the magnetic field. The first magnetic refrigeration device used a superconducting electromagnet \citep{Brown_1976}, and other systems also using a superconducting electromagnet have since been built \citep{Zimm_1998,Blumenfeld_2002,Rowe_2002}. Devices using a non superconducting electromagnet have also been constructed \citep{Bahl_2008,Coelho_2009}, but the greater majority of devices built in recent years have used permanent magnets to generate the magnetic field \citep{Bohigas_2000,Lee_2002,Lu_2005,Vasile_2006,Okamura_2007,Tura_2007,Zimm_2007,Zheng_2009,Engelbrecht_2009}.

The reason permanent magnets are preferred is that they do not require power to generate a magnetic field. This is not the case for an electromagnet where a large amount of power is needed to generate e.g. a 1 T magnetic flux density in a reasonable volume. This can be seen from the relation between the current, $I$, and the generated flux density, $B$, for an electromagnet in a single magnetic circuit consisting of a soft magnetic material with relative permeability, $\mu_\mathrm{r}$, and where the core has roughly the same cross sectional area throughout its length and the air gap is small compared with the cross sectional dimensions of the core,
\begin{eqnarray}
NI=B\left(\frac{L_\mathrm{core}}{\mu_\mathrm{r}\mu_0}+\frac{L_\mathrm{gap}}{\mu_0}\right)~,
\end{eqnarray}
where $N$ is the number of turns in the winding, $L_\mathrm{core}$ is the length of the soft magnetic material, $\mu_0$ is the permeability of free space and $L_\mathrm{gap}$ is the length of the air gap. In order to generate a 1.0 T magnetic flux density over e.g. a 30 mm air gap, which is typical for a magnetic refrigeration device, an iron cored solenoid with $\mu_\mathrm{r}=4000$ would need to have 24000 ampere windings. The length of the soft magnetic material is irrelevant as the expression is dominated by the second term. Such an electromagnet with 24000 ampere windings would need a massive power supply and an equally massive cooler to prevent the solenoid from overheating. Based on this simple calculation, it can be seen why an electromagnet is not preferred in most magnetic refrigeration devices.

A superconducting electromagnet is a better option than the traditional electromagnet because it requires little power to operate once the electromagnet has become superconducting as no power is lost to ohmic resistance. Although a superconducting electromagnet can create magnetic flux densities of the order of 10 T, continuous cooling is needed. This can be an expensive process and the apparatus surrounding the superconducting electromagnet can be of substantial size. However for large scale applications, e.g. large refrigerators for warehouses etc., a superconducting electromagnet might be a relevant solution. For common household refrigeration the superconducting electromagnet is at present not an option.

The only suitable choice left for generating the magnetic field is permanent magnets, which require no power to generate a flux density. The remainder of this paper will be concentrating on permanent magnet magnetic refrigerators, useable in common household refrigeration, as almost all research in magnetic refrigeration is focussed on this area. However the conclusions from this article will be applicable to any device using magnetocaloric materials, e.g. heat pumps, and not only magnetic refrigeration devices.

\section{Characterizing a magnet design}
When reviewing different magnet designs it is of the utmost importance that the different designs can be compared using a simple figure of merit. A previous suggestion for a comparison parameter was defined using the masses of the magnet and that of the magnetocaloric material used in the device \citep{Nikly_2007}. This parameter is not useful for two reasons: it contains no information about the magnetic flux density produced by the magnet design and using the same magnetic structure with two different magnetocaloric materials with different densities will yield different characterization results.

A general figure of merit, $M^{*}$, used to characterize a magnet design is defined by \citet{Jensen_1996} as
\begin{eqnarray}
M^{*}=\frac{\int_{V_\mathrm{field}}B^2dV}{\int_{V_\mathrm{mag}}B_\mathrm{rem}^2dV}
\end{eqnarray}
where $V_\mathrm{field}$ is the volume of the region where the magnetic field is created and $V_\mathrm{mag}$ is the volume of the magnets. It can be shown that the maximum value of $M^{*}$ is 0.25, and a structure is considered reasonably efficient if it has $M^{*} \geq 0.1$.

The strength of the magnetic field that is generated can also be quantified by a dimensionless number, $K$, which is the ratio between the magnetic flux density and the remanence of the magnets \citep{Coey_2003}. For a two dimensional structure with completely uniform remanence and magnetic flux density the two numbers $K$ and $M^{*}$ are related by the expression
\begin{eqnarray}
M^{*}=K^2\frac{A_\mathrm{field}}{A_\mathrm{mag}}~.
\end{eqnarray}
where $A_\mathrm{field}$ is the area of the high flux density region and $A_\mathrm{mag}$ is the area of the magnet. The figure of merit, $M^{*}$, often shown as a function of $K$, is useable for characterizing magnet designs in general, but for magnet design used in magnetic refrigeration the parameter does not take into account the flux density in the low field region of the magnet system where the magnetocaloric material is placed when it is demagnetized. Also, and more importantly, the scaling of the magnetocaloric effect itself with magnetic field is not taken into account. The importance of this will be considered shortly.

Finally a general performance metric for active magnetic refrigerators has been suggested \citep{Rowe_2009}. The cost and effectiveness of the magnet design is included in this metric as a linear function of the volume of the magnet. The generated flux density is also included in the metric. However, the metric does not make it possible to evaluate the efficiency of the magnet design alone.

Here the $\Lambda_\mathrm{cool}$ parameter proposed by \citet{Bjoerk_2008} will be used to characterize a magnet design for use in magnetic refrigeration. This parameter is designed to favor magnet designs that generate a high magnetic flux density in a large volume using a minimum of magnetic material. It also favors system designs in which the amount of time where the magnetic flux density is "wasted" by not magnetizing a magnetocaloric material is minimized.

\subsection{The $\Lambda_\mathrm{cool}$ parameter}
The $\Lambda_\mathrm{cool}$ parameter is a figure of merit that depends on a number of different parameters related to the magnetic assembly being evaluated.

\begin{figure}[!t]
  \centering
  \includegraphics[width=1.0\columnwidth]{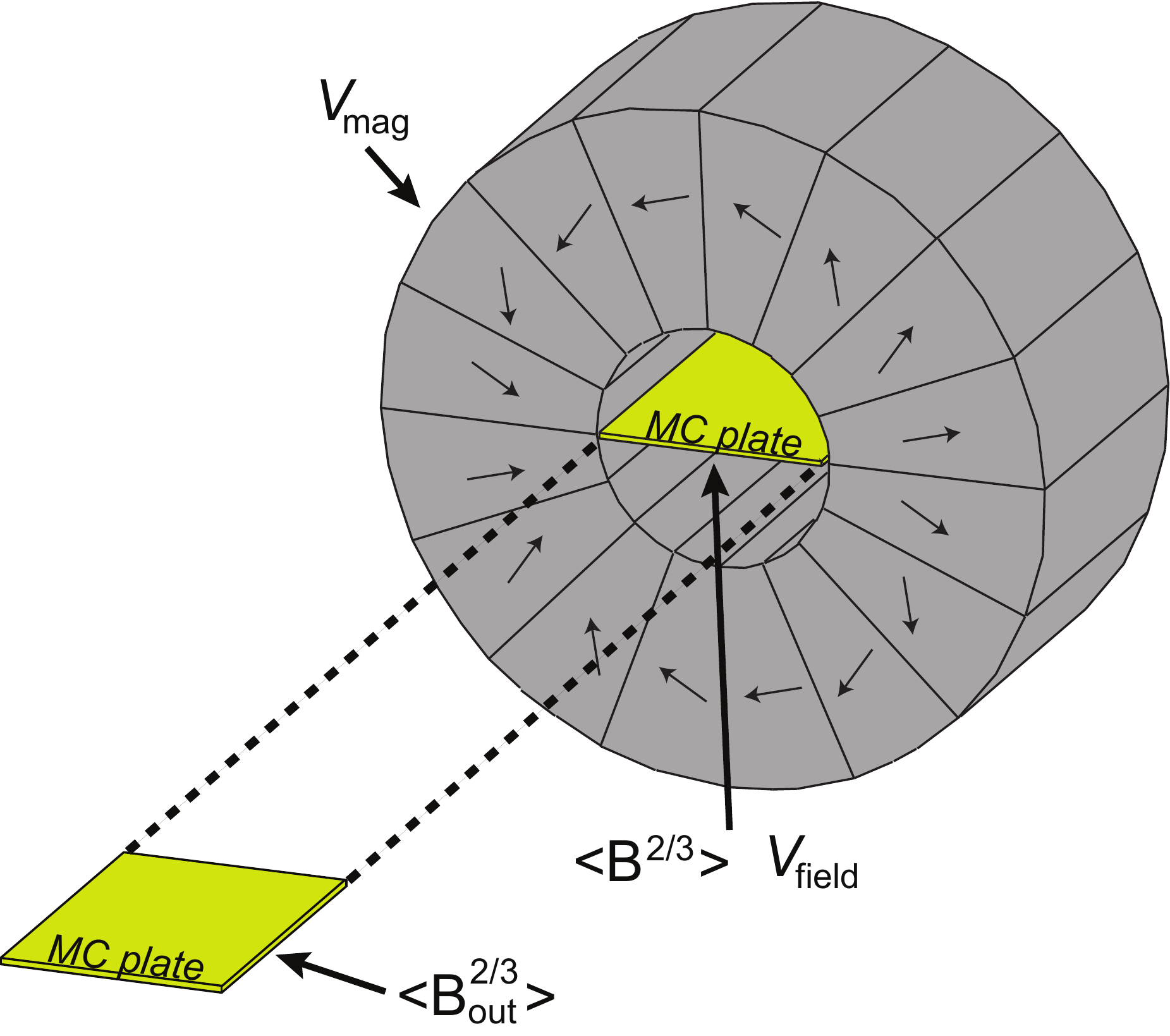}
  \caption{An illustration of some of the different variables in the $\Lambda_\mathrm{cool}$ parameter for the case of a Halbach cylinder. A plate of magnetocaloric (MC) material is shown in both the in and out of field position.}
  \label{Fig_Lambdacool_illustration}
\end{figure}

The $\Lambda_\mathrm{cool}$ parameter is defined as
\begin{eqnarray}
\Lambda_\mathrm{cool} \equiv \left(\langle B^{2/3}\rangle - \langle B^{2/3}_{\mathrm{out}}\rangle \right)\frac{V_{\mathrm{field}}}{V_{\mathrm{mag}}}P_{\mathrm{field}}~,
\end{eqnarray}
where $V_{\mathrm{mag}}$ is the volume of the magnet(s), $V_{\mathrm{field}}$ is the volume where a high flux density is generated, $P_{\mathrm{field}}$ is the fraction of an AMR cycle that magnetocaloric material is placed in the high flux density volume, $\langle B^{2/3}\rangle$ is the volume average of the flux density in the high flux density volume to the power of 2/3 and $\langle B^{2/3}_{\mathrm{out}}\rangle$ is the volume average of the flux density to the power of 2/3 in the volume where the magnetocaloric material is placed when it is being demagnetized. Some of these variables are illustrated for the case of a Halbach cylinder in Fig. \ref{Fig_Lambdacool_illustration}. Note that it is the magnetic flux density generated in an empty volume that is considered, and so $\mathbf{B}=\mu_0\mathbf{H}$, and thus it is equivalent to speak of the magnetic flux density or the magnetic field.

\begin{figure}[!t]
  \centering
\includegraphics[width=1.0\columnwidth]{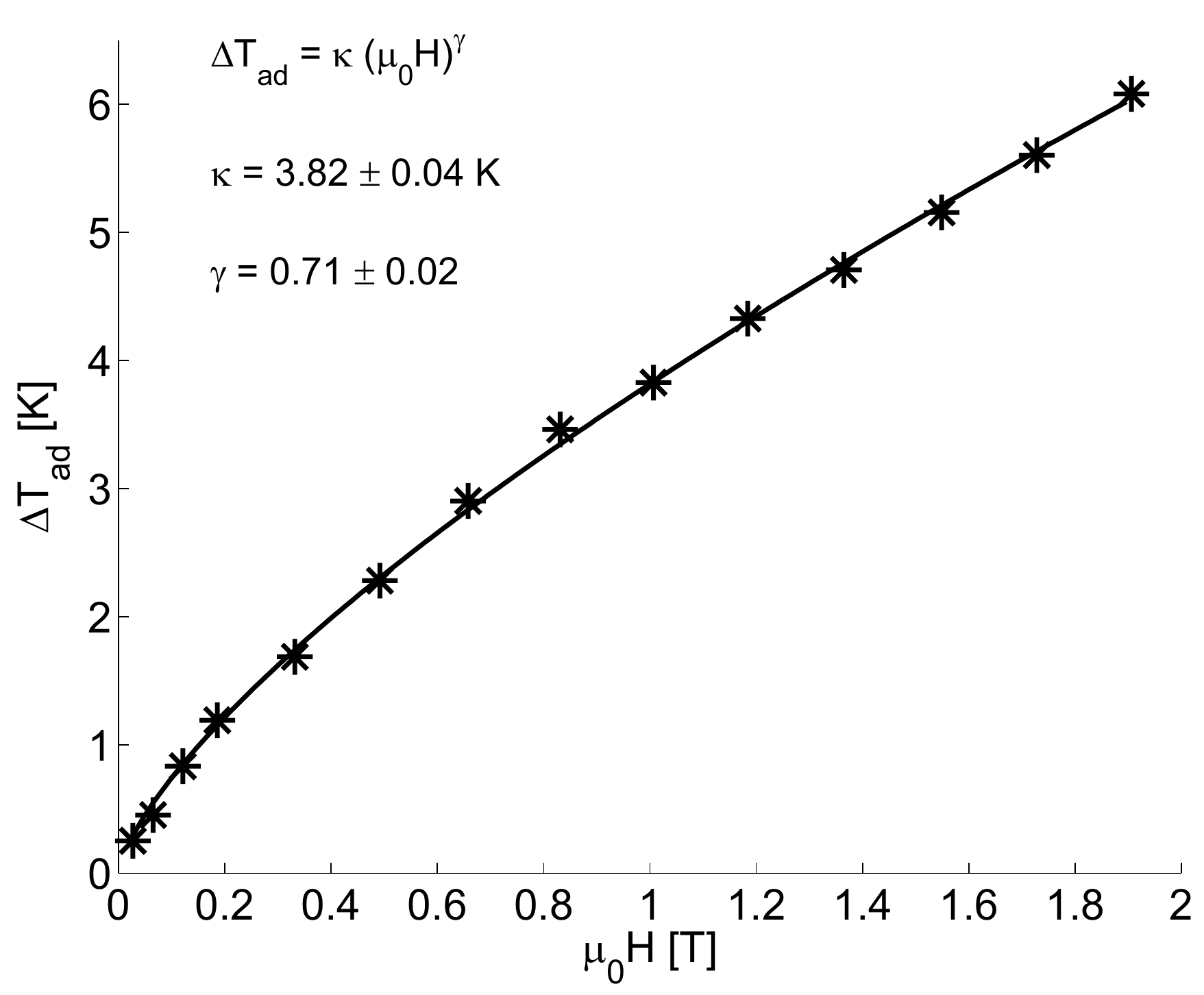}
  \caption{The scaling of the adiabatic temperature change of Gadolinium as a function of magnetic field at $T_\mathrm{c}$ ($293.6$ K). Data are from \citet{Pecharsky_2008} and are corrected for demagnetization using \citet{Aharoni_1998}.}
  \label{Fig_DeltaT_ad_scaling}
\end{figure}

Notice that $\Lambda_\mathrm{cool}$ depends on the flux density to the power of 2/3. The reason for this is that $\Lambda_\mathrm{cool}$ is defined to be proportional to the temperature change of the magnetocaloric material, and not the magnetic flux density, as the former is what is used to generate the temperature span and cooling power of the refrigeration device. This temperature change does not scale linearly with the magnetic flux density. A large number of different materials have been suggested as the active component of a magnetic refrigeration machine \citep{Gschneidner_2005}. The adiabatic temperature change at the Curie temperature of a general second order magnetocaloric phase transition material is predicted by mean field theory to scale with the power of 2/3 of the magnetic field \citep{Oesterreicher_1984}. This is in good accordance with the material most often used, i.e. the ``benchmark'' magnetocaloric material at room temperature, gadolinium, which has a magnetocaloric effect that scales with the magnetic field to the power of 0.7 at the Curie temperature \citep{Pecharsky_2006}, as also shown in Fig. \ref{Fig_DeltaT_ad_scaling}. This is why the $\Lambda_\mathrm{cool}$ parameter is proportional to the magnetic flux density to the power of 2/3. The scaling of the adiabatic temperature change away from $T_\mathrm{c}$ will in general be different from 2/3, but as long as the exponent is below 1 the conclusions of this article remain substantially unchanged. It should be noted that the entropy change of a number of magnetocaloric materials also scales as a power law with an exponent that in general is of the order of 2/3 \citep{Franco_2007, Dong_2008}.

It is not only the flux density in the magnetization region that is of importance to the magnetocaloric effect. The volume in which the magnetocaloric material is placed when it is demagnetized is equally important. In order to maximize the magnetocaloric effect, the flux density in this volume must be as low as possible. In a reciprocating device this can of course be accomplished by simply moving the magnetocaloric material far away from the magnet, but this will increase the physical size and cycle time of the magnetic refrigeration machine. In a rotating device the high and low flux density regions will generally be adjacent and care must be taken to minimize the ``leak'' of flux into the low flux density region.

To take into account the amount of magnetocaloric material that can experience a temperature change, the $\Lambda_\mathrm{cool}$ parameter is proportional to the volume of the high flux density region. Note that $\Lambda_\mathrm{cool}$ is proportional to the whole volume of the high flux density region and not only the volume occupied by the magnetocaloric material. Thus $\Lambda_\mathrm{cool}$ does not depend on the porosity of the magnetocaloric material, nor on the amount of e.g. plastic housing used to confine the magnetocaloric material. Also $\Lambda_\mathrm{cool}$ is inversely proportional to the volume of magnet material used, as the more magnet material used the more expensive the design will be.

Finally, the $\Lambda_\mathrm{cool}$ parameter is proportional to the fraction of the AMR cycle in which magnetocaloric material is placed in the high flux density volume. The reason for this is that if, e.g., magnetocaloric material is only placed inside the high flux density volume half the time of a full AMR cycle, the (expensive) magnet is not utilized during the remaining half of the cycle and it is thus essentially being wasted during this time. The fraction of time the magnetic flux generated by the magnet is being used to generate a magnetocaloric effect must be maximized.

One should note that the $\Lambda_\mathrm{cool}$ parameter will favor a design with a small magnetic flux density and large volume of the high flux density region. This is because the magnetic flux generated by a magnet scales with a power less than 2/3 with the volume of the magnet. In an actual device, heat transfer rates and thermal losses will set a lower limit on the flux density needed to produce a given temperature span and cooling capacity. Therefore for practical applications one would choose to optimize $\Lambda_\mathrm{cool}$ under the condition of a certain minimum flux density in the high flux density region.

The remanence of the magnets is not explicitly considered in the $\Lambda_\mathrm{cool}$ parameter. The reason for this is twofold. First this information is almost always not available for published magnet designs. Secondly the remanence of the NdFeB magnets used in all magnetic refrigeration magnet assemblies varies only between 1.2-1.4 T and so the exact value is not critical for comparison of different designs. Therefore, geometry accounts for almost all of the differences between different designs. Any soft magnetic material used in the magnet assembly is ignored, as the price of this material is in general much lower than that of the permanent magnets.

\section{Published magnet designs}
Having introduced the $\Lambda_\mathrm{cool}$ parameter, different published magnet designs can now be compared. There exist a substantial number of published designs of magnetic refrigerators but unfortunately many publications lack the necessary specifications to either reconstruct or directly calculate the $\Lambda_\mathrm{cool}$ parameter \citep{Richard_2004,Shir_2005,Zimm_2006,Buchelnikov_2007,Chen_2007,Vuarnoz_2007,Coelho_2009,Dupuis_2009,Sari_2009}. The designs presented below are the ones that represents the main magnets configurations and contain sufficient information to calculate $\Lambda_\mathrm{cool}$. A short description of each design is given prior to the calculation.

It should be noted that many of the magnetic refrigerators presented here are test devices and should be evaluated as such. However, it is also in the test design phase that large improvements to the design should be suggested. Therefore the evaluation of the designs can potentially lead to improvements for both current and future magnetic refrigerators.

For all designs an ``ideal'' device is considered when estimating the $P_{\mathrm{field}}$ parameter. In such a device the time for moving either the magnet or a bed of magnetocaloric material is minimized. This has been done in order that the $\Lambda_\mathrm{cool}$ parameter will not depend on, e.g., the power of the motor in the device. An example is the rotating design by \citet{Okamura_2007}, shown in a later section. Using the actual rotation speed of the magnet gives $P_{\mathrm{field}}=0.66$. However, we estimate that using a more powerful motor would allow $P_{\mathrm{field}}=0.9$. In the calculation of $\Lambda_\mathrm{cool}$ for the given design the latter value will be used. The AMR cycle is assumed to be symmetric, i.e. the magnetization and demagnetization steps are assumed to take the same amount of time.

The designs reviewed here have been classified into three groups, depending on the complexity of the design.
After all designs have been presented the designs are compared in Table \ref{Table_Dimensions}.

\subsection{Simple magnetic circuits}
The designs presented in this subsection all have a simple geometric structure and consist of rectangular blocks of magnets.

\subsubsection{Design by \citet{Zheng_2009}}
\begin{figure}[!t]
  \centering
\includegraphics[width=1.0\columnwidth]{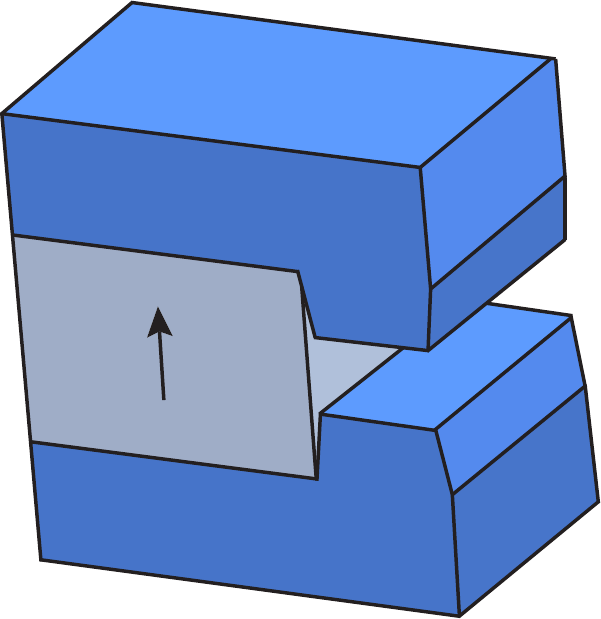}
  \caption{The design by \citet{Zheng_2009}. From \citet{Zheng_private_2009}. The arrow indicate the direction of magnetization of the magnet. The blue structure consists of soft magnetic material.}
  \label{Fig_Zheng}
\end{figure}

The general refrigerator design by \citet{Zheng_2009} is a reciprocating design where the magnet is moving  and two packed beds of magnetocaloric material are kept stationary. When one of the beds is in the magnetic field the other bed is out of the field. The flux density in the design is provided by a single rectangular magnet and the flux lines are guided by a soft magnetic material through a small air gap, as shown in Fig. \ref{Fig_Zheng}. Based on \citet{Zheng_private_2009} the volume of the magnet is 0.5 L and the volume of the high flux density region is 0.09 L. The mean magnetic flux density is 0.93 T.
Based on the cycle time, movement speed of the beds and the distance between these the actual $P_\mathrm{field}$ parameter is calculated to be 0.60. However using a faster and more powerful motor to move the magnet, as well as considering that the magnet has to be moved across a finite distance between the beds where no magnetocaloric material is present, the $P_\mathrm{field}$ parameter could be as high as 0.90.

\subsubsection{Design by \citet{Vasile_2006}}
\begin{figure}[!t]
  \centering
\includegraphics[width=1.0\columnwidth]{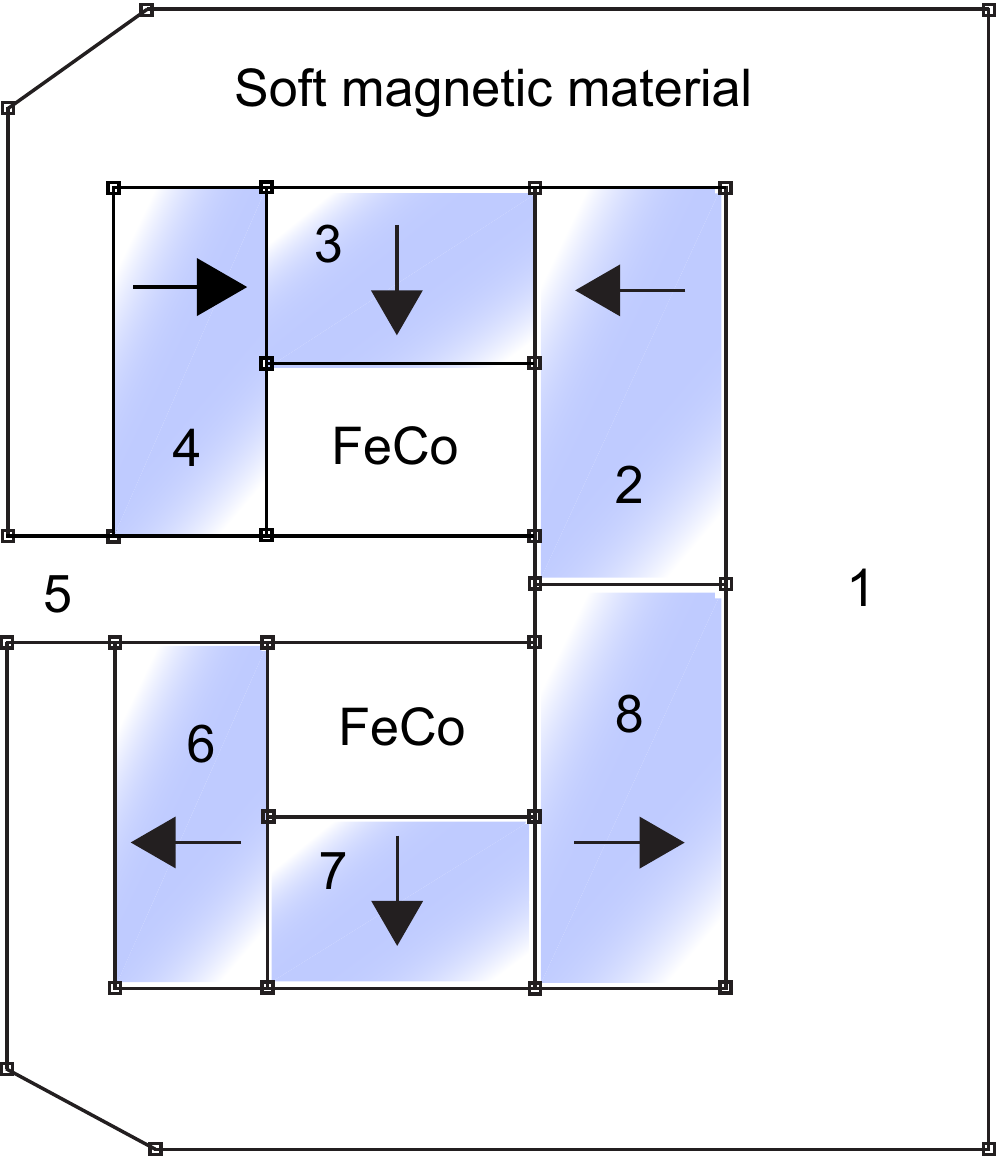}
  \caption{After \citet{Vasile_2006}. Reprinted with permission. (\copyright{}2006 Elsevier). The arrows indicate the direction of magnetization of the magnets.}
  \label{Fig_Vasile}
\end{figure}

The magnet design by \citet{Vasile_2006} is a ``C'' shaped magnet assembly of rectangular magnet blocks with soft magnetic material inside and outside of the ``C'' as seen in Fig. \ref{Fig_Vasile}. In this design the magnets are rotating around a circle with inserts filled with magnetocaloric material. The cross sectional area of the magnets is estimated to be 9.2 L/m and the high field gap cross sectional area to be 0.75 L/m. The magnetic flux density is given as 1.9 T in the high field region, but this is based on a two dimensional simulation so a real world assembly would have a significantly lower value. As the magnets are rotating continuously and the inserts for the magnetocaloric material fill most of the circle along which the magnet is rotating $P_\mathrm{field}$ is estimated to be 0.90.

\subsubsection{Design by \citet{Bohigas_2000}}
\begin{figure}[!t]
  \centering
\includegraphics[width=1.0\columnwidth]{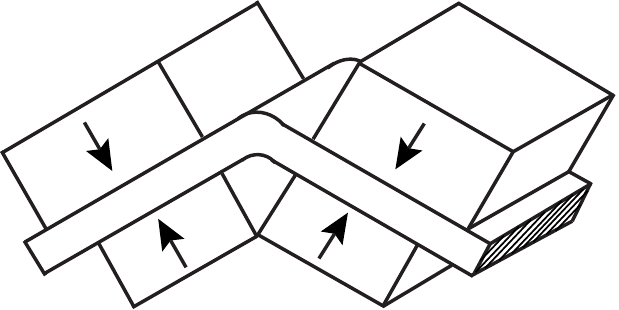}
  \caption{The design by \cite{Bohigas_2000}. Reprinted with permission. (\copyright{}2000 IEEE).}
  \label{Fig_Bohigas}
\end{figure}

The design by \citet{Bohigas_2000} is a rotating design in which the magnets are stationary and the magnetocaloric material is rotated in and out of the high flux density region. A total of eight rectangular magnets are used, four of them placed on the inside of the rotating wheel and four placed outside the wheel. The design can be seen in Fig. \ref{Fig_Bohigas}. The dimension of one of the inner blocks is given as 40$\times$40$\times$20 mm$^3$ and one of the outside blocks has dimensions 50$\times$50$\times$25 mm$^3$. The size of the air gap is given to be 7 mm and there are a total of four air gaps. From these figures we estimate the dimensions of one air gap to be 40$\times$7$\times$20 mm$^3$. Thus the volume of the magnets is 0.38 L and the volume of the high flux density region is 0.02 L. The flux density is given as 0.9 T. This design has magnetocaloric material continuously entering the high flux density region and thus the  $P_\mathrm{field}$ parameter is 1.

\subsubsection{Design by \citet{Tagliafico_2009}}
\begin{figure}[!t]
  \centering
\includegraphics[width=1.0\columnwidth]{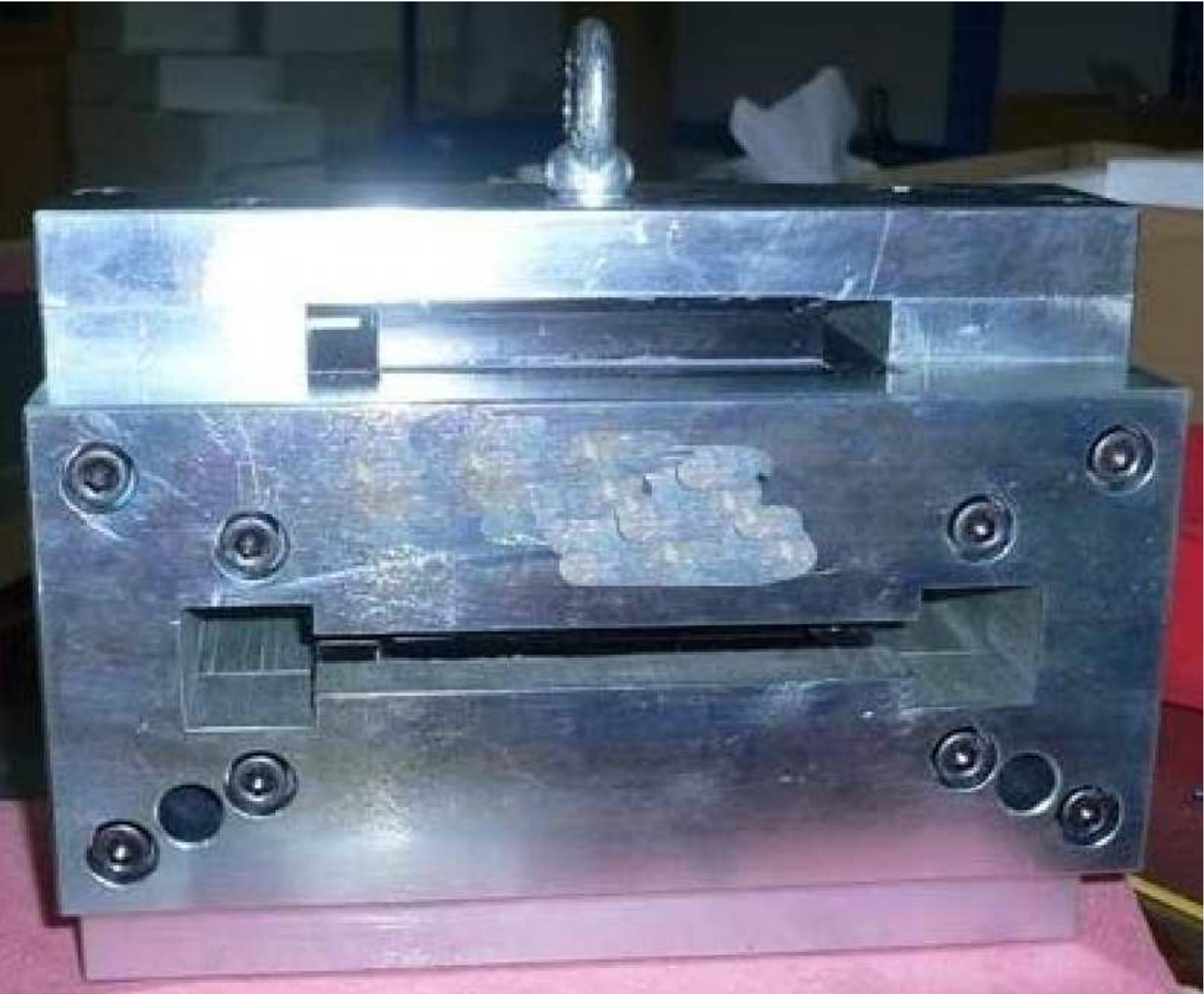}
  \caption{The magnet design by \citet{Tagliafico_2009} (\copyright{}2009 IIR/IIF). The magnetocaloric material passes through the gap in the structure.}
  \label{Fig_Tagliafico}
\end{figure}

The magnet design by \citet{Tagliafico_2009} consists of ten magnets in a rectangular structure which uses soft magnetic material to guide the flux lines round through the magnetic circuit. The magnet has a slot $50\times9.5\times100$ mm$^3$ in the center, through which the magnetocaloric material is moved, as seen in Fig. \ref{Fig_Tagliafico}. The volume of the high flux density region is thus 0.07 L. The flux density in the center of the slot is 1.55 T. A reported 5 kg of magnet is used, which corresponds to $V_\mathrm{mag}=0.68$ L. As two regenerative beds are run in parallel, and as the beds can be moved fairly quickly in and out of the high flux density region, the ideal $P_\mathrm{field}$ parameter is estimated to be $0.95$. The actual value for the $P_\mathrm{field}$ parameter, which can be estimated based on the total cycle time, is very close to this figure.

\subsubsection{Design by \citet{Tusek_2009}}
\begin{figure}[!t]
  \centering
\includegraphics[width=1.0\columnwidth]{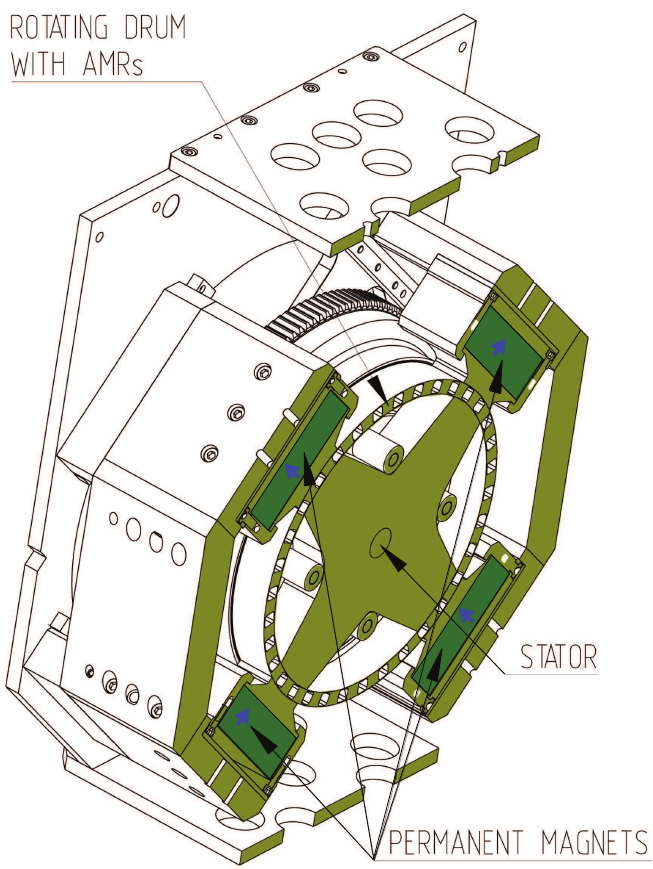}
  \caption{The design by \cite{Tusek_2009}. The magnetocaloric material is placed between the inner and outer magnetic structure. The direction of magnetization is shown as arrows on the magnet blocks. Adapted from \cite{Tusek_2009}.}
  \label{Fig_Tusek}
\end{figure}

The refrigeration system presented by \citet{Tusek_2009} uses a rotating AMR and a stationary magnet system. The magnet system consists of an inner and outer magnetic circuit with the magnetocaloric material placed in between the two structures. There are four high flux density regions and four low flux density regions along the circumference between the inner and the outer structure. A drawing of the design can be seen in Fig. \ref{Fig_Tusek}. The volume of the high flux density regions is four times $48\times10\times55$ mm$^3$, or 0.11 L. The amount of magnet material used is four times $90\times30\times90$ mm$^3$, or 0.65 L. The average mean flux density in the high field region is 0.97 T while it is 0.1 T in the low flux density region. The remanence of the magnets is 1.27 T. As magnetocaloric material is continuously rotated into the high field regions the magnets are constantly being used and thus $P_\mathrm{field}=1$.

\subsection{Halbach type magnet assemblies}
The magnetic structures presented in this subsection are all based on the Halbach cylinder design \citep{Halbach_1980,Mallinson_1973}.

\subsubsection{Design by \citet{Lee_2002}}
\begin{figure}[!t]
  \centering
\includegraphics[width=1.0\columnwidth]{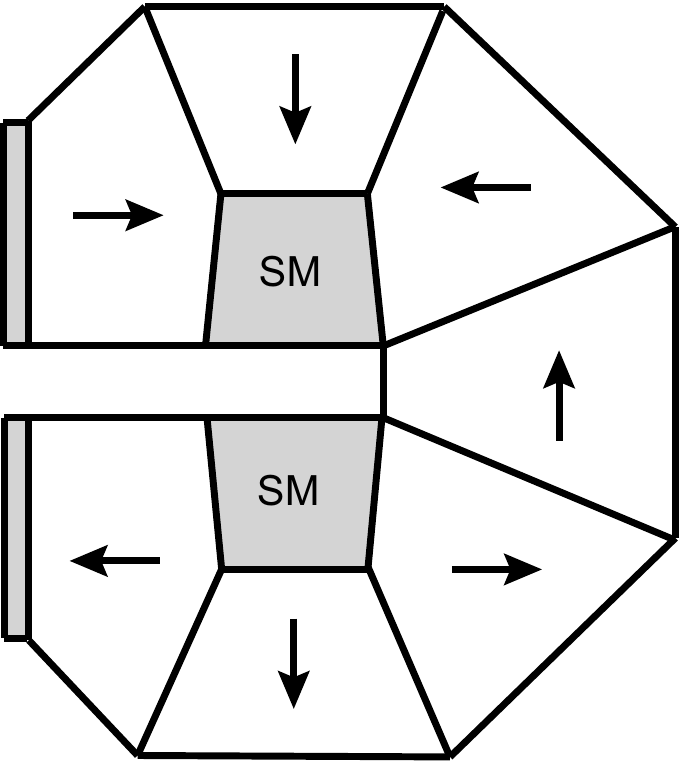}
  \caption{The design by \citet{Lee_2002}. The blocks labeled ``SM'' consists of soft magnetic material. Reprinted with permission. (\copyright2002 American Institute of Physics).}
  \label{Fig_Lee}
\end{figure}

The magnet design by \citet{Lee_2002} is suited to a reciprocating design with a stationary magnet and a moving bed of magnetocaloric material, but no actual device has been built. The magnet system is shaped like the letter ``C'', with a high homogenous flux density in the center. The design resembles an 8-segmented Halbach cylinder where one of the horizontal segmented has been removed. The flux density in the center is enhanced by blocks of soft magnetic material, placed in the center of the ``C''. An illustration of the design can be seen in Fig. \ref{Fig_Lee}. The design is very similar to the design by \citet{Vasile_2006} shown in Fig. \ref{Fig_Vasile}. However, this design is presented in this section because the shape of the magnets are more complex than in the latter design. The cross sectional dimensions of the array are given as 114$\times$128 mm$^2$ i.e. 14.6 L/m. The cross sectional area of the high flux region is estimated to be 25$\times$12.7 mm$^2$, i.e. 0.32 L/m. The magnetic flux density is given to be 1.9 T in the high flux region but this is based on a two dimensional simulation. Depending on the length of an actual device this figure will be significantly lower. No actual device has been built so the $P_\mathrm{field}$ is simply taken to be 0.90.

\subsubsection{Design by \citet{Engelbrecht_2009}}
\begin{figure}[!t]
  \centering
\includegraphics[width=1.0\columnwidth]{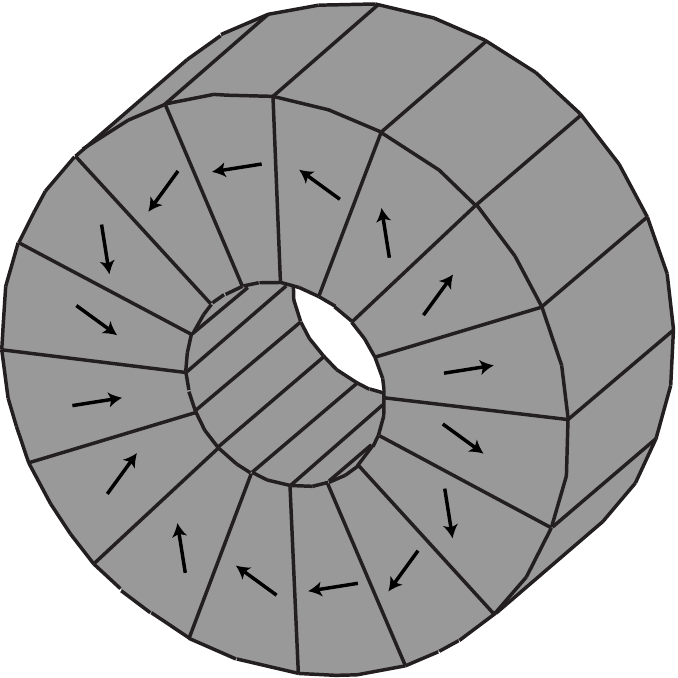}
  \caption{The design by \cite{Engelbrecht_2009}. The Halbach cylinder has an inner radius of 21 mm, an outer radius of 60 mm and a length of 50 mm.}
  \label{Fig_Engelbrecht}
\end{figure}

The magnetic refrigeration test machine designed at Ris\o{} DTU is a reciprocating device in which plates of magnetocaloric material are moved in and out of a stationary magnet \citep{Engelbrecht_2009}. The magnet is a Halbach cylinder consisting of 16 blocks of permanent magnets. The cylinder has an inner radius of 21 mm, an outer radius of 60 mm and a length of 50 mm. An illustration of the Halbach cylinder is shown in Fig. \ref{Fig_Engelbrecht}. The average magnetic flux density in the cylinder bore is 1.03 T. The volume of the magnet is 0.50 L and the volume of the high flux density region, i.e. the cylinder bore, is 0.07 L. The remanence of the magnets used in the Halbach cylinder is 1.4 T. The $P_\mathrm{field}$ parameter for this system design is 0.5. This is because for half the cycle time the stack of plates is out of the high field region leaving this empty. The actual $P_\mathrm{field}$ is slightly less than 0.5 due to the finite velocity of the moving regenerator.

\subsubsection{Design by \citet{Lu_2005}}
The magnetic refrigeration device designed by \citet{Lu_2005} is a reciprocating device with two separate packed beds of magnetocaloric material moving in and out of two stationary magnet assemblies to provide force compensation. Both magnets are 16 segmented Halbach cylinders with an inner radius of 15 mm and an outer radius of 70 mm. An illustration of the design is not shown as this design is very similar to the one shown in Fig. \ref{Fig_Engelbrecht}. The flux density produced is given as 1.4 T, and the length of the cylinder is 200 mm. Given these numbers the volume of the magnet is 2.94 L and the volume of the high flux density region is 0.14 L, for either of the magnets. For the same reasons as for the design by \citet{Engelbrecht_2009} the $P_\mathrm{field}$ parameter for this device is 0.5.

\subsubsection{Design by \citet{Kim_2009}}
The magnet design by \citet{Kim_2009} is a 16 segmented Halbach cylinder. A single bed of magnetocaloric material is reciprocated through the cylinder bore. The radius of the cylinder bore is 8 mm, the outer radius of the cylinder is 38 mm and the length is 47 mm. An illustration of the design is not shown as this design is very similar to the one shown in Fig. \ref{Fig_Engelbrecht}. The volume of the high flux density region is 0.01 L while the volume of the magnet is 0.20 L. The flux density is 1.58 T at the center of the bore and 1 T at the edge, with a mean value of 1.4 T. As only a single magnetocaloric bed is used the high flux density region is only used half the time, and thus $P_\mathrm{field}$ is 0.5.

\subsubsection{Design by \citet{Tura_2007}}
\begin{figure}[!t]
  \centering
\includegraphics[width=1.0\columnwidth]{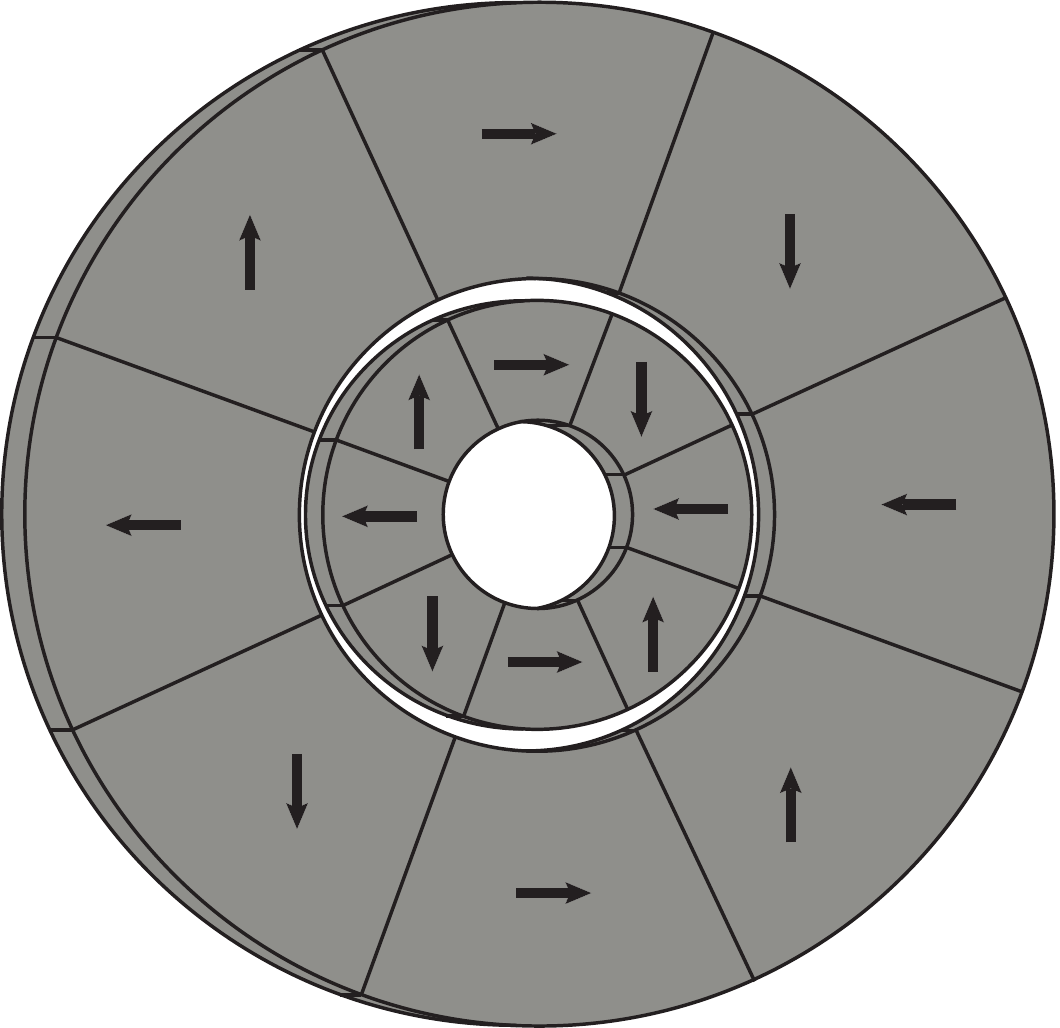}
  \caption{A sketch of the concentric Halbach magnet design by \citet{Tura_2007}, viewed from the front. The inner and outer radius of the inner cylinder is 12.5 mm and 27 mm respectively while the corresponding figures for the outer cylinder is 30 mm and 60 mm respectively. The length of the actual concentric cylinder is 100 mm. The rotational configuration shown here is the high flux density configuration.}
  \label{Fig_Tura}
\end{figure}

The magnetic refrigerator presented by \citet{Tura_2007} is a rotating system in which the magnetocaloric material is kept stationary and a magnet is rotated to alter the flux density. An illustration of the design can be seen in Fig. \ref{Fig_Tura}. The magnet design used in the device consists of two separate magnets each of which consists of two concentric Halbach cylinders. The reason that two separate magnets are used is that the system can be run such that the magnetic forces are balanced.
In the concentric Halbach cylinder design the flux density in the inner cylinder bore can be controlled by rotating the inner or outer magnet. \citet{Tura_2007} report that when the inner magnet is rotated the mean magnetic flux produced can be changed continuously from 0.1 T to 1.4 T. The total volume of the magnetic material is 1.03 L, while the total volume of the high flux density region is 0.05 L \citep{Rowe_private_2009}. These values are for one of the concentric Halbach cylinders. The remanence of the blocks in the inner cylinder is 1.15 T while for the outer magnet it is 1.25 T. The  $P_\mathrm{field}$ parameter for this system design is 0.5 as half of a cycle the inner magnet will be turned such that it cancels the magnetic flux generated by the outer magnet. In this configuration there is no high flux density region, and the magnets are not being used to generate cooling.

\subsection{Complex magnetic structures}
The designs presented in this subsection have a complex structure and consists of irregularly shaped magnet blocks.

\subsubsection{Design by \citet{Zimm_2007}}
\begin{figure}[!t]
  \centering
\includegraphics[width=1.0\columnwidth]{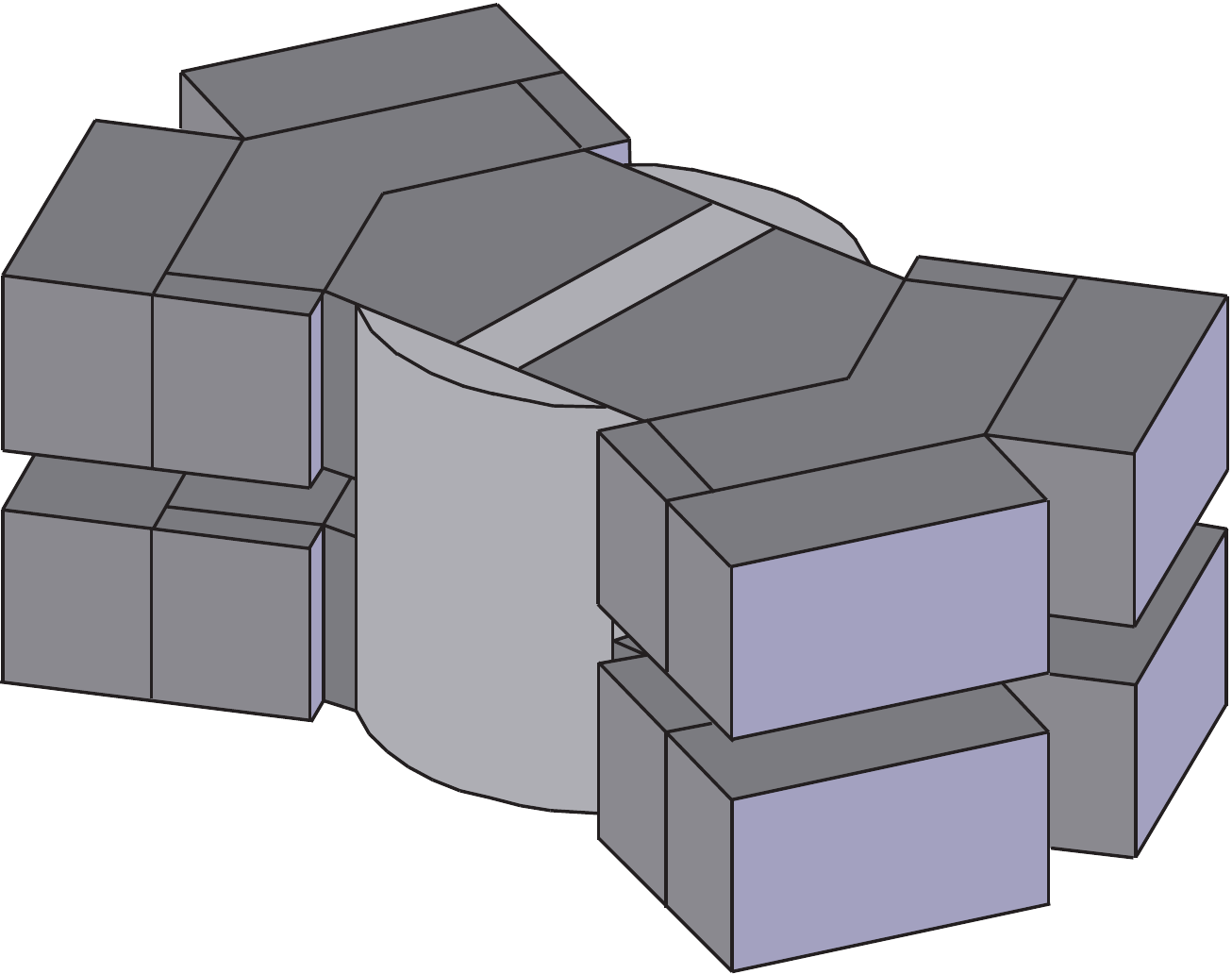}
  \caption{The complex magnet design by \citet{Zimm_2007} (\copyright{}2007 IIR/IIF). The magnetocaloric material passes through the gap between the upper and lower ``Y'' structures. The dark grey blocks are individual magnets, while the light grey structure is made of soft magnetic material. The direction of magnetization of the individual blocks are taken from \citet{Chell_2006}.}
  \label{Fig_Zimm}
\end{figure}

The magnetic refrigeration machine presented by \citet{Zimm_2007} utilizes a rotating design in which the magnetocaloric material is stationary and the magnet is rotating. The magnet design is quite complex, utilizing both magnets and soft magnetic materials, but essentially consists of two Y-shaped magnetic structures separated by an air gap. The design is shown in Fig. \ref{Fig_Zimm}. The high flux density region spans an angle of 60 degrees on two opposite sides of the design. Based on \cite{Chell_private_2009} the total volume of the magnet assembly is 4.70 L, the volume of the high flux density region is 0.15 L and the mean flux density is 1.5 T. The  $P_\mathrm{field}$ parameter for this design is essentially given by the speed at which the magnet rotates from one bed of magnetocaloric material to the next. These are separated by an angle of 30 degrees. If the magnet is rotated fast the $P_\mathrm{field}$ parameter could be as high as 0.90.

\subsubsection{Design by \citet{Okamura_2007}}
\begin{figure}[!t]
  \centering
\includegraphics[width=1.0\columnwidth]{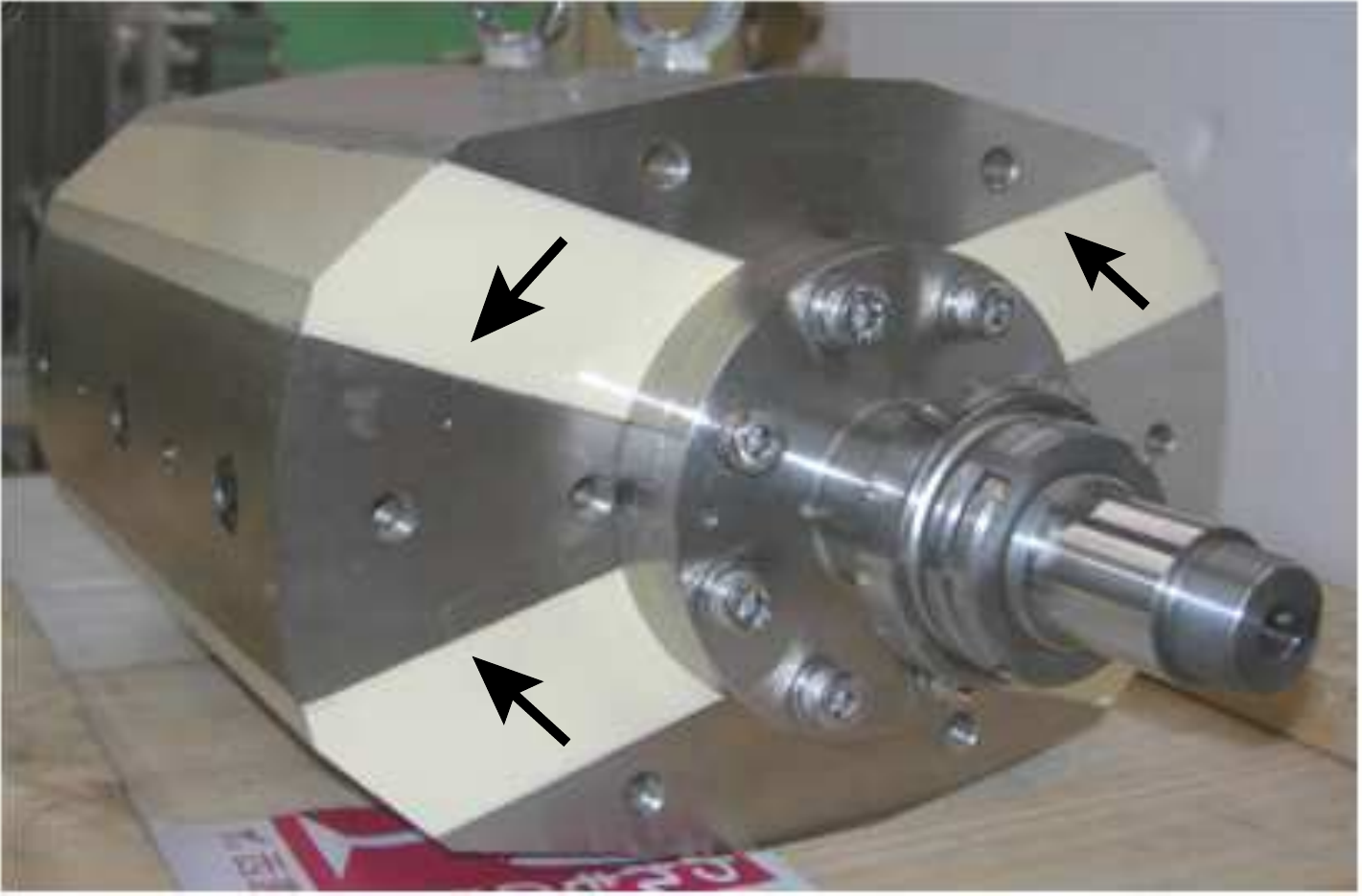}
  \caption{The inner magnetic structure in the design by \cite{Okamura_2007}. From \cite{Okamura_private_2009}. The outer magnetic structure consists of a cylinder of soft magnetic material (not shown). The arrows indicate the direction of magnetization of the magnets, which are white in color.}
  \label{Fig_Okamura}
\end{figure}

The design by \citet{Okamura_2007} is a rotating device in which the magnet is rotated past ducts packed with magnetocaloric material. The magnet design consists of a complex arrangement of permanent magnets and soft magnetic materials which is assembled in the shape of an inner rotor consisting both of magnets and soft magnetic material with an outer yoke consisting of only soft magnetic material. The magnetocaloric material is placed in four ducts in the air gap between the inner and outer structure. The inner rotor is designed such that magnets with identical poles are facing each other and separated by a soft magnetic material. This increases the flux density and "pushes" the flux lines from the inner rotor to the outer yoke. A photo of the design can be seen in Fig. \ref{Fig_Okamura}. The mean flux density is 1.0 T and the magnet design contains 3.38 L of magnet and 0.80 L of high flux density region \citep{Okamura_private_2009}. As with the design by \citet{Zimm_2007} the  $P_\mathrm{field}$ parameter for this design is essentially given by the speed at which the magnet rotates from one duct to the next. The actual $P_\mathrm{field}$ parameter can be estimated using the total cycle time and the time to rotate between two ducts, separated by an angle of 40 degrees, and is found to be 0.66. However a faster rotation might be possible and thus we estimate that the $P_\mathrm{field}$ parameter can be as high as 0.90.

\section{Comparing the designs}
In Table \ref{Table_Dimensions} the different magnet designs are presented. In the table the $\Lambda_\mathrm{cool}$ parameter has been calculated for each design, thus allowing a direct comparison of the designs.

\begin{table*}[!t]
\centering
\caption{
The specifications of different magnet designs used in magnetic refrigeration devices. In all cases is it assumed that $\langle B \rangle ^{2/3} = \langle B^{2/3} \rangle$, which is only true if the flux density is completely homogenous. $^*$ designates a quantity estimated by the authors of this article. $^\mathrm{H}$ indicates that the value of the flux density is the highest possible attainable flux density in the center of the design, and as such is not a representative average of the magnetic flux density for the whole of the high flux density region. $^\mathrm{2D}$ indicates that the flux density is based on a two dimensional simulation. These notoriously overestimate the flux density except for very long assemblies and so $\Lambda_\mathrm{cool}$ will be overestimated for these designs. Some of the two dimensional designs also have their volumes given per meter.}
\label{Table_Dimensions}
\begin{tabular}{lcccccp{4.5cm}cc}
Name & $V_{\mathrm{mag}}$ & $V_{\mathrm{field}}$ & $\langle B \rangle$ & $\langle B_{\mathrm{out}} \rangle$ & $P_{\mathrm{field}}$ & Magnet type & $\frac{\Lambda_{\mathrm{cool}}}{P_{\mathrm{field}}}$ & $\Lambda_{\mathrm{cool}}$\\
     & [L] & [L] & [T] & [T] & & & & \\\hline
\citet{Bohigas_2000}     & 0.38   & 0.02   & 0.9$^\mathrm{H}$      & 0$^*$   & 1        & Rectangular magnets on round surface    & 0.05  & 0.05 \\
\citet{Engelbrecht_2009} & 0.5    & 0.07   & 1.03                  & 0       & 0.5      & Halbach cylinder                        & 0.14  & 0.07  \\
\citet{Kim_2009}         & 0.20   & 0.01   & 1.4                   & 0       & 0.5      & Halbach cylinder                        & 0.06  & 0.03  \\
\citet{Lee_2002}         & 14.6/m & 0.32/m & 1.9$^\mathrm{H, 2D}$  & 0$^*$   & 0.90$^*$ & ``C'' shaped Halbach cylinder           & 0.03  & 0.03  \\
\citet{Lu_2005}          & 2.94   & 0.14   & 1.4$^\mathrm{H}$      & 0       & 0.5      & Halbach cylinder                        & 0.06  & 0.03  \\
\citet{Okamura_2007}     & 3.38   & 0.80   & 1.0                   & 0       & 0.90$^*$ & Inner magnet rotor, soft magnetic yoke  & 0.24  & 0.21  \\
\citet{Tagliafico_2009}  & 0.68   & 0.07   & 1.55$^\mathrm{H}$     & 0       & 0.95     & Rectangular magnetic circuit with slot  & 0.14  & 0.13  \\
\citet{Tura_2007}        & 1.03   & 0.05   & 1.4                   & 0.1     & 0.5      & Concentric Halbach cylinders            & 0.05  & 0.03  \\
\citet{Tusek_2009}       & 0.11   & 0.65   & 0.97                  & 0.1     & 1        & Stationary magnet, rotating MC material & 0.13  & 0.13  \\
\citet{Vasile_2006}      & 9.2/m  & 0.75/m & 1.9$^\mathrm{H, 2D}$  & 0$^*$   & 0.90$^*$ & ``C'' shaped circuit                    & 0.12  & 0.11  \\
\citet{Zheng_2009}       & 0.5    & 0.09   & 0.93                  & 0$^*$   & 0.90$^*$ & Single magnet magnetic circuit          & 0.17  & 0.15  \\
\citet{Zimm_2007}        & 4.70   & 0.15   & 1.5                   & 0.1$^*$ & 0.90$^*$ & ``Y'' shaped magnetic structure         & 0.04  & 0.03  
\end{tabular}
\end{table*}

\begin{figure*}
\begin{center}
\includegraphics[width=1.1\textwidth]{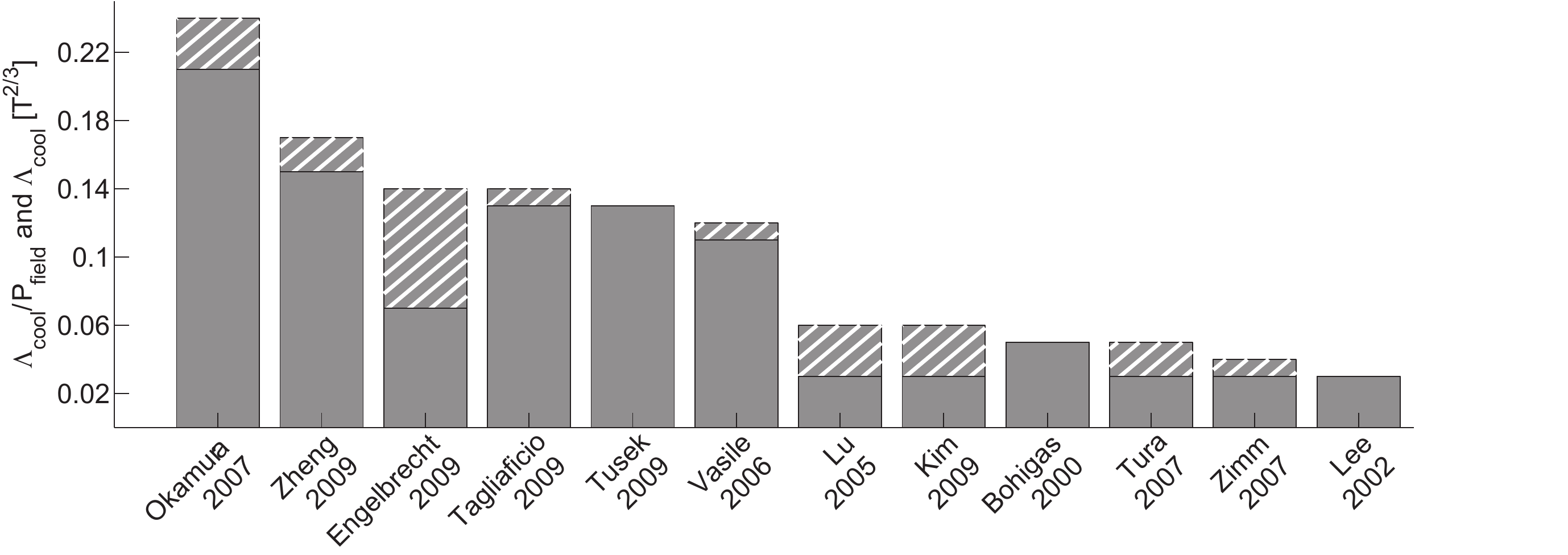}
\caption{The parameters $\Lambda_{\mathrm{cool}}/P_{\mathrm{field}}$ (hatched) and $\Lambda_{\mathrm{cool}}$ (full). The $\Lambda_{\mathrm{cool}}/P_{\mathrm{field}}$ parameter only takes the magnet design into account and not the fraction of a cycle the magnet is used. As $P_{\mathrm{field}} \leq 1$ the $\Lambda_{\mathrm{cool}}$ parameter is always less than or equal $\Lambda_{\mathrm{cool}}/P_{\mathrm{field}}$. Completely filled bars have $P_{\mathrm{field}} = 1$. Note that the best design is five times as good as the design with the lowest value of $\Lambda_{\mathrm{cool}}/P_{\mathrm{field}}$.}
\label{Fig_Lambda_cool_Combined}
\end{center}
\end{figure*}

In Fig. \ref{Fig_Lambda_cool_Combined} the parameter $\Lambda_\mathrm{cool}/P_{\mathrm{field}}$, which only takes the magnet assembly into account and not the design of the refrigeration device, as well as the actual $\Lambda_\mathrm{cool}$ parameter are shown. From the figure it is seen that the magnet design by \citet{Okamura_2007} outperforms the remaining magnet designs. Compared to \citet{Lu_2005} the design by \citet{Okamura_2007} uses almost the same amount of magnets but creates a high flux density region over three times larger. An interesting thing to note is that although the design by \citet{Zimm_2007} creates a very high flux density the design has a rather low $\Lambda_\mathrm{cool}$ value because the magnetocaloric temperature change only scales with the magnetic field to the power of 2/3 at the Curie temperature and this, as mentioned previously, does not favor high flux densities. However $\Lambda_\mathrm{cool}$ should be optimized under the condition of a certain minimum flux density in the high flux density region, e.g. the flux density required to obtain a given temperature span of the device. It is also seen that many of the reciprocating designs only utilize the magnet in half of the AMR cycle, i.e. that the $P_\mathrm{field}$ parameter is 0.5. This means that the expensive magnet is only utilized half the time, which is very inefficient. It is also seen that the different Halbach cylinders do not perform equally well. This is because the efficiency of a Halbach cylinder is strongly dependent on the relative dimensions of the cylinder \citep{Bjoerk_2008}.

Note that the actual magnetic refrigeration machines, when ranked by their temperature span and cooling capacity, does not necessarily follow the trend of Fig. \ref{Fig_Lambda_cool_Combined} \citep{Engelbrecht_2007,Gschneidner_2008,Rowe_2009}. This can be caused by different types of magnetocaloric material, different regenerator designs and different operating parameters.

Having evaluated existing magnet designs we now analyze the advantages of these designs and focus on how to design the optimal magnet for a magnetic refrigerator. The optimal design is limited by the energy density in the magnets themselves. Also for, e.g., very large Halbach cylinders the coercivity of the magnet is a limiting factor because the magnetic field is opposite to the direction of magnetization around the inner equator of the Halbach cylinder \citep{Bloch_1998,Bjoerk_2008}. A standard grade NdFeB magnet with a remanence of 1.2 T has a intrinsic coercivity of $\mu_0 H_\mathrm{C} = 3.2$ T, so the reversal of the magnet will only be a problem above this flux density. One should note that for NdFeB magnets with a higher energy density, e.g. 1.4 T, the intrinsic coercivity can be significantly lower, e.g. around $\mu_0 H_\mathrm{C} = 1.4$ T.

\subsection{Design of an optimal magnet assembly}
Based on the knowledge gained from the magnet assemblies reviewed certain key features that the magnet assembly must accomplish or provide can be stated. It must produce a region that has a high flux density preferably with as high uniformity as possible. Also the magnet must be designed such that the amount of leakage of flux or stray field is as low as possible. This includes both leakage to the surroundings and leakage to low flux density regions in the magnet assembly. The recommendations to maximize $\Lambda_\mathrm{cool}$ for a given flux density can be summed up as
\flushleft
\begin{itemize}
\item Use minimum amount of magnets
\item Make the volume for magnetocaloric material as large as possible
\item Utilize the magnet at all times
\item Ensure that the flux density in the low flux density region is low
\item Minimize leakage to surrounding by e.g. using soft magnetic material as flux guides
\item Use the lowest possible flux density necessary to obtain the chosen temperature span and cooling capacity
\end{itemize}

If magnetic refrigeration is to become a viable alternative to conventional refrigeration technology these simple design criteria must be followed.

\section{Conclusion}
Different ways of generating the magnetic field used in a magnetic refrigeration device have been discussed and it has been shown that permanent magnets are the only viable solution, at present, to common household magnetic refrigeration devices. Twelve published magnet designs were reviewed in detail and were compared using the $\Lambda_\mathrm{cool}$ parameter. The best design was found to be five times better than the worst design. Finally guidelines for designing an optimal magnet assembly was presented.

\section*{Acknowledgements}
The authors would like to acknowledge the support of the Programme Commission on Energy and Environment (EnMi) (Contract No. 2104-06-0032) which is part of the Danish Council for Strategic Research. The authors also wish to thank T. Okamura, A. Rowe, C. Zimm, J. Chell, Z.G. Zheng and J. Tu\v{s}ek for useful discussions and for providing some of the figures and values in this article.

\bibliographystyle{elsarticle-harv}

\end{document}